\documentclass[12pt]{article}
\usepackage[hmargin=2cm,bottom=3cm]{geometry}
\usepackage{amsmath}
\usepackage{amssymb}
\usepackage{graphicx,color}
\usepackage[font=scriptsize]{caption}
\usepackage{empheq}
\usepackage{listings}
\usepackage{hyperref}
\usepackage[affil-it]{authblk}
\usepackage{lineno}
\usepackage[round]{natbib}

\makeatletter
\def\@maketitle{%
  \newpage
  \null
  \vskip 2em%
  \begin{center}%
  \let \footnote \thanks
    {\Large\bfseries \@title \par}%
    \vskip 1.5em%
    {\normalsize
      \lineskip .5em%
      \begin{tabular}[t]{c}%
        \@author
      \end{tabular}\par}%
    \vskip 1em%
    {\normalsize \@date}%
  \end{center}%
  \par
  \vskip 1.5em}
\makeatother

\newcommand*{\defeq}{\mathrel{\vcenter{\baselineskip0.5ex \lineskiplimit0pt
                     \hbox{\scriptsize.}\hbox{\scriptsize.}}}%
                     =}

\providecommand{\e}[1]{\ensuremath{\cdot 10^{#1}}}

\date{}
\title{Characterization of the Endemic Equilibrium and Response to Mutant Injection in a Multi-strain Disease Model}
\author[1,2]{Tom\'{a}s Aquino}
\author[1]{Diogo Bolster}
\author[2]{Ana Nunes\thanks{anunes@ptmat.fc.ul.pt}}
\affil[1]{Department of Civil \& Environmental Engineering and Earth Sciences\\
	University of Notre Dame\\
	46556 Indiana, USA}
\affil[2]{Centro de F\'{i}sica da Mat\'{e}ria Condensada and Departamento de F\'{i}sica\\
Faculdade de Ci\^{e}ncias da Universidade de Lisboa\\
P-1649-003 Lisboa Codex, Portugal}

\begin{document}

\maketitle
\abstract{Starting from common assumptions, we build a rate equation model for multi-strain disease dynamics in terms of immune repertoire classes. We then move to a strain-level description where a low-order closure reminiscent of a pair approximation can be applied. We characterize the endemic equilibrium of the ensuing model in the absence of mutation and discuss the presence of degeneracy regarding the prevalence of the different strains. Finally we study the behavior of the system under the injection of mutant strains.}

\section{Introduction}
\label{S::Introduction}

Many human pathogens exhibit antigenic diversity, with several strains that elicit different immune responses coexisting
in the population or within individual hosts at a given time, or replacing one another over time. Infectious
diseases of this class still defy effective control, and improving our understanding of their dynamics is of great practical importance. Modeling multi-strain infections is in itself a theoretical challenge, because typical compartmental models of mathematical epidemiology keep track of the infection histories of the hosts \citep{chavez:89, andreasen:97}. This results in a number of variables that increases exponentially with the number of strains, making the problem intractable even for a relatively modest number of strains competing through cross-immunity. Ferguson and co-workers \citep{ferguson:2004, ferguson:2005, ferguson:2008} characterized the symmetric equilibrium of a fixed, arbitrary number of strains with identical epidemiological parameters, but it has proven very difficult to go beyond the description of equilibrium properties in this framework. 

Two strategies have been proposed to reduce the complexity of the problem using reasonable simplifying assumptions.
Gog and co-workers \citep{gog:2002a, gog:2002b} introduced the idea of a status based approach,
where different compartments in the population are associated with current immunity profiles rather than with the individuals' infection history. In this approach, each host is either susceptible or immune to any given strain $i$ (polarized immunity). Partial immunity conferred by infection with another strain $j$ translates into a fraction of those infected by $j$ becoming immune to $i$, rather than all the infected becoming partially immune. They showed that this approach, combined with the assumption of 'reduced infectivity', results in a drastic reduction of the number of variables needed to describe the system. Reduced infectivity means that an immune host exposed to a given strain will not transmit that strain, but will develop an immune response just as a susceptible infected with that strain would. By contrast with 'reduced susceptibility', whereby immunity prevents infection and there is no change in the host's immune repertoire, under reduced infectivity immunity prevents further transmission but does not prevent infection. While this assumption may be difficult to sustain from the biological point of view, it makes the scaling of the number of variables of the system with the number of strains linear, instead of exponential, allowing for the exploration of large sets of competing strains.

The second approach, put forward in \citep{levin:2007}, is also status based. It takes as dynamic variables the fraction of the population that is immune to and the fraction that is infected with each strain. The equations that determine the time evolution of the fraction of immune individuals involve, apart from the cross-immunity properties of the set of strains, the fraction of the population that is immune to pairs of strains. In order to close the system, time evolution equations for these must be derived, which in turn involve the fraction of the population that is immune to strain triplets, and so on. In much the same way as is usually done for spatially extended systems, this process can be truncated at a given order via a moment closure, i.e., an ansatz that expresses n-tuplets in terms of the lower order variables. This elegant idea works without the disputable reduced infectivity assumption. It allows for the reduction of the number of variables that describe the system, which becomes polynomial in the number of strains, the actual order depending on the order of the closure. 

A benchmark for a successful theoretical model is its ability to reproduce the antigenic drift and shift typical of influenza A when pathogen mutation is included. Both \citep{gog:2002b} and \citep{levin:2007} offer applications to influenza A and show that in strain spaces of dimension 1 and 2 the models do indeed recover the essential qualitative features of influenza evolution, namely the steady strain replacement over the years combined with limited diversity at any given time. 
In higher dimensional strain spaces, however, extinction or explosive diversity are the generic
outcomes of the simplest models \citep{girvan:2002, tria:2005, ballesteros:2009, koelle:2009}. Theoretical proposals 
relying on specific assumptions regarding strain cross-immunity were shown to generate patterns of antigenic evolution compatible with influenza A \citep{ferguson:2003, koelle:2006, koelle:2009, bedford:2012}. However, their actual implementation as computational models  involves a combination of features and is too complex to provide a generally accepted modeling framework for multi-strain infection dynamics. Indeed, agent-based simulations give us the insight of virtual experiments but,
even with modern computational facilities, they can be too time consuming to enable a full exploration of parameter space. Additionally, they
are not amenable to analytic treatment.
 In recent years, along with other contributions in agent-based modeling frameworks \citep{cobey:2011, buckee:2011, zinder:2013}, the effort to deal with multi-strain dynamics in a deterministic setting has been continued \citep{adams:2007, adams:2009, minayev:2009a, minayev:2009b, koelle:2010, kucharski:2012}.

In this paper we add to this effort, using the approach of \citep{levin:2007} as a starting point, so as to avoid the large number of degrees of freedom that comes with infection-history-based rate equation models. 
We model each strain as a set of integers that represent different configurations of the pathogen's antigenic sites or epitopes and we stipulate an immune response in the host population with some degree of heterogeneity. Different hosts may build antibodies to different numbers and different sets of antigenic sites. Together with the assumption that a single matching antibody provides lifelong immunity to a strain, the immune response scheme completely determines the cross-immunity structure of the strain set.

In Section~\ref{S::Model} we set up the mathematical model, and in Section~\ref{S::Endemic_eq} we study the endemic equilibrium, adding to the results of \citep{ferguson:2004, ferguson:2005, ferguson:2008}. In  Section~\ref{S::Injection} we consider a single founding strain in endemic equilibrium and investigate the possible outcomes, coexistence, substitution or extinction, of an immunity evading mutation, following the system until a new equilibrium is reached. To deal with extinction and mutation in a description based on continuous variables that represent population densities, we define a cut-off, the inverse of the effective population size, such that if densities become lower than this cut-off they are taken to be zero. A related study was presented in  \citep{adams:2007,adams:2009}, where the fate of the mutant strain was assessed only from the initial value of the time derivative of the corresponding density of infected.

\section{Model}
\label{S::Model}

In this model we assume a simple setup where each viral strain is characterized by a certain number of epitopes, each of which is in one of a number of possible configurations. All epitopes have equal roles, and the infectious properties of viral strains are all identical; their effectiveness at infecting a host is fully determined by the host's immune history. Upon infection with a strain, the host will produce a certain number of antibodies, each of which matches the antigene configuration at a particular epitope. In general, the immune system builds a polyclonal response upon infection with a strain, that is, it will produce antibodies that respond to all or a large fraction of the virus's epitopes. However, there is evidence of variation in the individual immune responses to identical strains and it has been proposed that children's immune system may have a monoclonal response \citep{nakajima:2000,sato:2004}, meaning that with each infection a child will build an antibody for only one of the epitopes.
In the proposed framework, we model different responses by a probability that an individual will produce a certain number of antibodies.  With these assumptions in mind, let us derive the rate equation formulation of the model for multi-strain competition in a well mixed population.

A given strain is characterized by a set of $n_e$ integers in $\{ 0, \ldots, n_c \}$, each representing a different configuration of the antigene at a certain epitope. This comprises a total of $N_s = n_c^{n_e}$ strains. At any particular time, an individual may be infected with at most a single strain; they are said to be susceptible if they are not infected with any strain. The assumption of no co-infections (no infection by more than one strain) is arguable, but can be defended in two ways: (i) a sick individual tends to stay home, and thus isolate himself from contact with new infections, and (ii) upon infection the immune system is highly active and responds more effectively to secondary infections.

Additionally, each individual in the population has an immune history made up of $n_e$ sets of integers, representing antibodies against specific configurations of the antigenes at each of the $n_e$ epitopes. To make the model amenable to the analytical treatment that will be presented in subsection~\ref{subS::strain-level}, we further assume that having one matching antibody for any of the epitope configurations is enough to grant total immunity against a strain. In formal terms, we say an individual with immune history $A = \{A_1,...,A_{n_e}\}$ is immune to a strain $i = \{i_1,...,i_{n_e}\}$ if $\exists j : i_j \in A_j$, and in this case we write $i \in A$. Otherwise, we write $i \notin A$.

Consider now the formalization of immunity acquisition. Take an individual with immune history $A$ as discussed above. Upon infection with strain $i$, the immune repertoire becomes $B = \{B_1,...,B_j\}$, where for each $j$ we have either $B_j = A_j$, if an antibody was not produced for epitope $j$, or $B_j = A_j \cup \{i_j\}$, if an antibody was produced for that epitope. For each infection, consider that the number of antibodies produced follows a distribution $p_{ \{ 1 \leqslant \alpha \leqslant n_e \} }$, where $p_\alpha$ represents the probability of $\alpha$ antibodies being produced. Furthermore, consider that antibodies for each epitope have the same probability of being produced, leaving aside the possibility of epitope immunodominance \citep{cobey:2011}, which for a given $p_{\{\alpha\}}$ would reduce the diversity of immune responses.

Let us now define the processes that describe the dynamics. We assume that the population is at demographic equilibrium, that is, the rate of birth of individuals equals the rate of death. An important assumption made at this point is that all individuals have the same death rate, meaning we do not consider an age structure or disease-related death. Finally, we consider a well mixed population, or in other words pairwise interactions occur with the same probability between any two individuals in appropriate states. Let $I^k$ be the fraction of individuals infected with strain $k$ and $S$ be the fraction of susceptible (not infected) individuals. The following processes are then present:

\begin{enumerate}
\item \textbf{Birth-death:} At rate $\mu$, each individual becomes na\"{i}ve (immune repertoire $A = \varnothing$) and susceptible.
\item \textbf{Infection:} At rate $\beta S$, each individual infected with strain $k$ tries to infect a susceptible. The susceptible becomes infected if it is not immune to $k$. The overall rate of attempted infection per individual in the population is $\beta I^k S$.
\item \textbf{Recovery:} At rate $\gamma$, each infected individual becomes susceptible. The overall rate per individual is $\gamma I$.
\item \textbf{Acquiring Immunity:} Infected individuals are immune to all strains.  Upon recovery, an individual adds $\alpha$  antibodies corresponding to the infecting strain. The number $\alpha > 1$ follows the prescribed distribution $p_{\{\alpha\}}$. Each possible antibody is produced with the same probability.
\item \textbf{Mutation:} At rate $m$, each strain present in the population changes a random antigene to a random new one, if the infected individual is not immune to the new strain. We say the old strain mutated to the new one. The overall rate of attempted mutation per individual is $m I$.
\end{enumerate}

\subsection{Immune history description}

This model is a generalization of the simple classical Susceptible-Infected-Recovered (SIR) model to a system with multiple circulating strains. To develop equations for our system, we need to set up classes corresponding to each immune history, representing different immunity profiles in the population due to contact with different diseases. We also need to specify how these classes evolve into one another, that is, how immunity is acquired upon infection. As we will see, this characterizes the cross-immunity profile, that is, the way strains confer immunity to each other. Finally, we must specify the structure of ``strain space" by defining what strains are allowed to mutate into one another.

Let $S_A$ be the fraction of individuals not infected with any strain, generically referred to as ``healthy'' or ``susceptibles", with immune history $A$, and $I^i_A$ be the fraction of individuals infected with strain $i$ with immune history $A$. Let $C(A,k,B)$ denote the probability that upon infection with strain $k$ the immune history of an individual changes from $B$ to $A$. Define also $M_i$ as the mutational neighborhood of $i$, i.e., the set of strains to which $i$ may mutate (here corresponding to strains that are related by changing the configuration of a single epitope).

Then the dynamics presented above may be described in a rate equation formalism as:

\begin{subequations}
\label{eq::rate_history}
\begin{empheq}[left=\empheqlbrace]{align} 
	\label{eq::rate_history_1}
	\dot{S}_A &= \mu \left(\delta_{A,\varnothing} - S_A \right)
		+ \gamma \sum_{ k } I^k_A
		- \sum_{ k } \left(1 -  \delta_{k,A} \right) \beta I^k S_A \; ,
	\\
	\label{eq::rate_history_2}
	\begin{split}
	\dot{I}^j_A &= \sum_{ B } \sum_{ k } C( A, k, B )\left( 1 -  \delta_{j,B} \right)
		\left(
			\delta_{jk} \beta I^k S_B
			+ m_{jk} I^k_B
		\right) \; +\\
		&\;\;\; - ( m + \gamma + \mu ) I^j_A \; ,
	\end{split}
\end{empheq}
\end{subequations}
where $m_{jk} = \delta_{j,M_k} \, m / |M_k|$. Here and throughout, $|.|$ applied to a set denotes the cardinality, or number of elements. We use the Kronecker Delta symbol in a generalized manner: if $i$,$j$ are both strains, then $\delta_{ij} = 1$ if $i=j$, and zero otherwise. Similarly, if A and B are both immune history sets, $\delta_{A,B}$ equals one if $A=B$, and zero otherwise. Finally, if $i$ is a strain and A is an immune history or mutational neighborhood, $\delta_{i,A} = 1$ if $i \in A$, and zero otherwise.

The first set of equations describes the time evolution of the fraction of susceptibles with immune history $A$. The first term describes death of susceptibles with any immune history and birth of na\"{i}ve (immune history $A=\varnothing$) susceptibles. The second term represents the increase of susceptibles with immune history $A$ due to recovery of individuals infected with any strain and immune history $A$. The last term represents the decrease of susceptibles with immune history $A$ due to infection with each strain $k$; the $\delta_{k,A}$ accounts for the fact that susceptibles immune to $k$ (that is, for which $k \in A$) cannot be infected by strain $k$.

The second set of equations describes the time evolution of the fraction of individuals infected with strain $j$ and immune history $A$. The first term has two components, corresponding to each of the terms in the final parenthesis. The first term in the parenthesis accounts for infection with strain $j$ of susceptibles with immune history $B$. The second accounts for mutation of the strain $k$ infecting individuals with immune history $B$ to strain $j$. Both these terms are multiplied by $C(A,k,B)$ and summed over all $B$ and $k$; this accounts for acquiring of immunity that leads from history $B$ to history $A$. The Kronecker Delta $\delta_{jk}$ is introduced simply for notational compactness and ensures that only strain $j$ contributes to that term. The $1-\delta_{j,B}$ factor deals with the fact that individuals immune to $j$ cannot be infected by that strain, whether through contact or through mutation. Finally, the last term in this equation accounts for the decrease in individuals infected with strain $j$ due to mutation, recovery, and natural death.

There are $2^{N_s} ( N_s + 1  )$ equations in this description. However, they imply one conservation condition that arises from demographic equilibrium and says that the total number of individuals in the population is constant:

\begin{equation}
	\sum_A \sum_k I^k_A + \sum_A S_A = I + S = 1 \; .
\end{equation}
Thus, we have $2^{N_s} ( N_s + 1  ) - 1$ independent equations.

\subsection{Strain-level description}
\label{subS::strain-level}

As mentioned above, this type of rate-equation formalism rapidly leads to a very large number of equations. This is because we must keep track of all possible immune histories; if there are $N_s$ strains circulating, there are $2^{N_s}$ different immune histories, corresponding to an individual having or not having an antibody corresponding to each particular epitope configuration. Correspondingly we find on the order of $2^{N_s}$ equations, as we saw above. This leads to two main complications: (i) it is difficult to deal with the enormous number of ensuing equations, either analytically or even numerically, and (ii) the very large number of different immune histories in relation to the population size means that we subdivide the population into a large number of classes which correspondingly have very small numbers of elements. This can make stochastic effects important, putting into question the deterministic rate equation formalism for reasonable population sizes. Note that the rate equation formalism implicitly considers the limit of infinite population sizes by considering fractions of individuals as continuous variables.

To simplify system \eqref{eq::rate_history}, we make use of the general strategy outlined in \citep{levin:2007}
which requires that: (i) immunity is not lost from an infection process, (ii) each strain confers total immunity to itself, and (iii) acquiring of immunity is independent of previous history. All these are satisfied by the model described here: (i) is trivial, and (ii) and (iii) are a consequence of the fact that in this model at least one antibody is produced upon infection and one antibody is enough to grant immunity. Adapting the technique of \citep{levin:2007} to this model, we define the immunity variables:

\begin{equation}
	\xi_i = \sum_{ A : i } S_A \; , \;\;\; \eta^j_i = \sum_{ A : i } I^j_A \; ,
\end{equation}
where $A : i$ is shorthand for $A : i \in A$. The physical meaning of these variables should be clear: $\xi_i$ represents the fraction of healthy individuals immune to strain $i$, and $\eta_i^j$ represent the fraction of infected with strain $j$ that are immune to strain $i$. Note that, because we have chosen to account for immunity updating at the time of infection, being infected with strain $i$ requires being immune to it, and so $\eta^i_i$ represents the fraction of individuals infected with strain $i$.
The idea here is to describe the system in terms of a set of variables that allows for  reasonable approximations. Although it is easier to write the equations for the system in the form given by~\eqref{eq::rate_history}, taking the point of view of the single strain as opposed to generic immune history will lead to natural approximations on the way strains interact through cross-immunity.

Define $C_i(k,B) = \sum_{ A : i } C(A,k,B)$, the probability of ending up with immunity to strain $i$ upon infection with strain $k$. In the model described above we have, for $i \notin B$,
$C_i(k,B) \equiv \sigma_{ik} \equiv \sum_{\alpha=1}^{n_e} p_\alpha \sigma^{(\alpha)}_{ik}$, where $\sigma^{(\alpha)}_{ik}$ corresponds to the production of exactly $\alpha$ antibodies, which happens with probability $p_\alpha$. Simple combinatorial arguments tell us that:

\begin{equation}
	\sigma^{(\alpha)}_{ik} = \sigma^{(\alpha)}_{ki} = 
\begin{cases}
	1 \; ,  &| i \cap k | > n_e - \alpha \; ,\\
	1 - \frac{ ( n_e -  | i \cap k | ) ! }{ ( n_e -  | i \cap k | - \alpha ) ! } \frac{ (n_e - \alpha)! }{ n_e! }  \; ,
		& | i \cap k | \le n_e - \alpha \; .
\end{cases}
\label{eq::cases}
\end{equation}
Here, $i \cap k$ denotes the set of matching antigenes between strains $i$ and $k$. We find that $\sigma_{ik}$ is independent of previous immune history $B$. On the other hand, $C_i(k,B) = 1$ if $i \in B$, that is, immunity is not lost upon infection. The matrix $(\sigma_{ik})$ is central to the model and encodes the cross-immunity granted by each strain to every other.

Expressions~\eqref{eq::cases} become particularly simple for $\alpha =1$ and for $\alpha = n_e$. For $\alpha =1$, \eqref{eq::cases} reads:

\begin{equation}
	\sigma^{(1)}_{ik} = \sigma^{(1)}_{ki} =  \frac{| i \cap k |}{n_e}\; ,
\end{equation}
meaning that cross-immunity of two strains is simply given by their fraction of identical epitopes. For $\alpha = n_e$,
\begin{equation}
	\sigma^{(n_e)}_{ik} = \sigma^{(n_e)}_{ki} = 
\begin{cases}
	1 \; ,  &| i \cap k | > 0\; ,\\
	0 \; ,
		& | i \cap k | = 0\; ,
\end{cases}
\end{equation}
which simply states that each strain confers total cross-immunity to any other strain so long as it shares at least one antigene with it, and there is no cross-immunity between completely different strains.

Appropriately summing over immune histories in equations~\eqref{eq::rate_history}, we find:

\begin{subequations}
\label{eq::strains}
\begin{empheq}[left=\empheqlbrace]{align}
	\label{eq::strains::xi}
	\dot{\xi}_i &= \sum_k \left[ 
		\gamma \eta^k_i - \beta \eta^k_k( \xi_i - \xi_{ik})
		\right] - \mu \xi_i \; ,
	\\
	\label{eq::strains::eta}
	\begin{split}
	\dot{\eta}^j_i &= \beta \eta_j^j
				\left[
					\xi_i - \xi_{ij} + \sigma_{ij} ( 1 - I  - \xi_i - \xi_j + \xi_{ij})
				\right] +\\
			&\;\;\; + \sum_k m_{jk}
				\left[ \eta^k_i - \eta^k_{ij} + \sigma_{ij} ( I^k  - \eta^k_i - \eta^k_j + \eta^k_{ij} )
				\right] +\\
			&\;\;\; - ( m + \gamma + \mu ) \eta^j_i \; .
	\end{split}
\end{empheq}
\end{subequations}
Here, $\xi_{ij}$ and $\eta^{k}_{ij}$ are so-called second-order variables and denote fractions of the population that are immune to $i$ and $j$ simultaneously, while being respectively healthy or infected with strain $k$. Note that this description would be exact if we knew the time evolution of these variables. However, describing their evolution requires introducing further equations, which will depend on further variables describing simultaneous immunity to higher numbers of multiple strains. This in turn will require the introduction of equations for higher order terms. Thus, to make the problem tractable, we must truncate the system at some order by specifying a so-called closure relation, a heuristic description of a higher-order variable describing simultaneous immunity in terms of lower-order variables. Here we adopt a closure of order one, that is, we close the system by providing descriptions for the second order variables. The main strength of this approach is that it becomes easier to make assumptions based on physical ideas about the structure of the cross-immunity interactions between strains, leading to a significant simplification of the original system. We introduce a particular closure below. With a closure of this order, there are a total of $N_s( N_s + 1 )$ equations in this description. This can still be a large number, but represents a considerable reduction from exponential to quadratic in the number of strains. 

We note that a simple and reasonable closure for $\xi_{ij}$ is \citep{levin:2007}:

\begin{equation}
\label{eq::closure}
	\xi_{ij} = (1 - \sigma_{ij} )\xi_i \xi_j + \sigma_{ij} \text{min}( \xi_i, \xi_j ) \; .
\end{equation}
For each pair $i,j$ of strains, this expression interpolates between the $\sigma_{ij} = 0$ scenario, where the probability of being immune to $i$ and $j$ are independent, and the $\sigma_{ij} = 1$ scenario, where infection by $i$ guarantees immunity to $j$.

In order to bring equations~\eqref{eq::strains} to closed form when $m \neq 0$, a similar assumption should be made to express $\eta ^k_{ij}$ in terms of $\eta ^k_{i}$, $\eta ^k_{j}$. However, since we will side step \eqref{eq::strains} when dealing with mutations, this additional closure assumption will be left unspecified. The set~\eqref{eq::strains} of ordinary differential equations (ODEs) with $m=0$ together with the closure assumption~\eqref{eq::closure} describes the behaviour of our model in the absence of mutations and will be called from now on the ODE model.

\section{Endemic equilibrium}
\label{S::Endemic_eq}

Let us now look at the endemic equilibrium of equations~\eqref{eq::strains} in the absence of mutations.

\subsection{Deriving an equation for equilibrium}

Equations~\eqref{eq::strains::eta} for $i=j$, that is, for the total fractions of infected with each strain, tell us that, for all $j$:

\begin{equation}
	\eta_j^j \left[
			R_0 ( 1 - I - \xi_j ) - 1
		     \right] = 0 \; ,
\end{equation}
where
$R_0 =  \beta /( \gamma + \mu )$. Now let 
$\Lambda = \{ j : \eta_j^j \neq 0 \}$.
Then we have $I = \sum_{j \in \Lambda} \eta^j_j$, and for all circulating strains $j \in \Lambda$ we find:

\begin{equation}
\label{eq::sym_xi}
	\xi_j = \xi = 1 -  I - R_0^{-1}\; .
\end{equation}
This shows that the existence of an endemic equilibrium requires $R_0 >  1$, which is directly analogous to results from classical SIR models. Because all strains in this model are equivalent as they have the same $\beta$ and $\gamma$, the equilibrium also forces all immunity variables $\xi_j$ corresponding to the circulating strains to take the same value.

In equilibrium, defining
$\eta_i = \sum_j \eta_i^j$,
equations~\eqref{eq::strains::xi} for $i \in \Lambda$  can be written as:

\begin{equation}
\label{eq::etai::1}
	 - \gamma \eta_i -
			\beta \sum_{ j \in \Lambda } \eta_j^j \xi_{ij} +
			\beta I \xi + \mu \xi 
			= 0 \; .
\end{equation}
Summing over all $j$ in equations~\eqref{eq::strains::eta} leads to

\begin{equation}
\label{eq::etai::2}
	\eta_i = R_0
		\left[
			\xi I  +
			\sum_{ j \in \Lambda } \eta_j^j
			\left[
				- \xi_{ij} +
				\sigma_{ij} ( 1 - I - 2 \xi  + \xi_{ij} )
			\right]
		\right] \; .
\end{equation}
Substituting this result into eq.~\eqref{eq::etai::1} and using \eqref{eq::sym_xi} for $I$, we find

\begin{equation}
\label{eq::etas}
	\sum_{ j \in \Lambda } C_{ij} \eta_j^j = \frac{ \mu }{ \gamma } \xi ( 1 - \xi ) \; ,
\end{equation}
where

\begin{equation}
\label{eq::cij}
	C_{ij}  = \xi_{ij} 
		\left[ 
			\sigma_{ij} + \frac{\mu}{\gamma} 
		\right] +
		\sigma_{ij} \left( R_0^{-1} - \xi \right) \; .
\end{equation}

Equation \eqref{eq::etas} is a linear system of $\tilde N_S$ equations in $\tilde N_S$ variables, where $\tilde N_S = | \Lambda |$ is the number of circulating strains. 
Because for a given closure 
$\xi_{ij} = f( \xi_i, \xi_j, \sigma_{ij} )$,
$\xi_{ij}$ becomes a function of $\xi$ and $\sigma_{ij}$ for 
$i, j \in \Lambda$;
the solution of eq.~\eqref{eq::etas} gives the individual prevalences as a function of $\xi$, which can in turn be found by substitution into eq.~\eqref{eq::sym_xi}. For a certain solution for the $\eta_j^j$, the crossed $\eta_i^j$ can be found through the equilibrium of eq.~\eqref{eq::strains::eta}:

\begin{equation}
\label{eq::cross}
	\eta_i^j = R_0 \eta_j^j
			\left[
				\xi_i - \xi_{ij} + \sigma_{ij} ( 1 - I  - \xi_i - \xi_j + \xi_{ij})
			\right] \; .
\end{equation}
In particular, note that for all $i$ we have $ j \notin \Lambda \Rightarrow \eta_i^j = 0$.

\subsection{Symmetric equilibrium}

Let $\delta_j \defeq \eta^j_j - I / |\Lambda|$. Note that as a consequence $\sum_{j \in \Lambda} \delta_j = 0$. Substituting into eq.~\eqref{eq::etas}, we find

\begin{equation}
\label{eq::symm_cond}
	\sum_{ j \in \Lambda } C_{ij} \delta_j
		+ \frac{I}{|\Lambda|}\sum_{ j \in \Lambda } C_{ij} = 
			\frac{ \mu }{ \gamma } \xi ( 1 - \xi ) \; .\\
\end{equation}

For a symmetric equilibrium to exist, $(\delta_j)_{j \in \Lambda} = 0$ must be a solution. In that case, we must have

\begin{equation}
\label{eq::iind_cond}
	\sum_{ j \in \Lambda } C_{ij} = 
			\frac{|\Lambda|}{I} \frac{ \mu }{ \gamma } \xi ( 1 - \xi ) \; .\\
\end{equation}
In particular, $\sum_{ j \in \Lambda } C_{ij}$ must be independent of $i$.
Conversely, if $\sum_{ j \in \Lambda } C_{ij}$ is independent of $i$ then 
eq.~\eqref{eq::iind_cond} together with eq.~\eqref{eq::sym_xi} become an equation for the total infective density $I$ whose form will depend on the closure assumption. 
Given a solution $I$ of this equation, the individual prevalences are then $\eta_j^j = I / | \Lambda |$ for all $j$.
Assuming that there is at least one solution $I$ in the interval $(0,1]$, we conclude that the system has symmetric equilibria if and only if $\sum_{ j \in \Lambda } C_{ij}$ is independent of $i$. Since by construction the lines of $(\sigma_{ij})$ are permutations of each other (all strains are equivalent), and since we have found that $C_{ij}$ depends on $i$ and $j$ only through $\sigma_{ij}$, this condition is always fulfilled when $\Lambda$ coincides with the set of all strains.

Let us then  study 
eq.~\eqref{eq::iind_cond} for the total prevalence in a symmetric equilibrium with all strains present.
Define
$\bar{\sigma} = \sum_{i \in \Lambda} \sigma_{ij} / | \Lambda |$, for any $j$.
This is well defined because, since the lines of the matrix $(\sigma_{ij})$ are permutations of each other, this quantity is independent of $j$. For the same reason, we may define the $j$-independent (equilibrium) quantities
$\bar{s} = \sum_{i \in \Lambda} \xi_{ij} / | \Lambda |$
and
$\tilde s = \sum_{i \in \Lambda} \sigma_{ij} \xi_{ij} / | \Lambda |$.
With these definitions, by summing over all $i$ in eq.~\eqref{eq::etas} we obtain an equation for the total infected density $I$:

\begin{equation}
\label{eq::I_raw}
	 I
	 \left[ 
	 	\bar{ \sigma }
		\left( R_0^{-1}
			  - \xi  
		\right) +
		\frac{\mu}{\gamma} \bar{s}
		 +
		\tilde s
	\right] =
		\frac{ \mu }{ \gamma }
			\xi ( 1 -\xi ) \; ,
\end{equation}
where $\xi$, $\bar{s}$ and $\tilde s$ are to be understood as functions of $I$ due to the equilibrium constraints eq.~\eqref{eq::sym_xi}. 

To proceed, we must now settle for a specific closure relation. Let us consider the one defined by eq.~\eqref{eq::closure}. Using~\eqref{eq::sym_xi} for $\xi$, some algebra leads to an explicit cubic equation for $I$ depending only on system parameters:

\begin{equation}
\label{eq::I}
\begin{split}
	P_3(I) = &\, R_0 \left[
		\bar{\sigma} ( 1 - \tilde\sigma ) +
		\frac{ \mu }{ \gamma }( 1 - \bar{\sigma} )
	\right] I^3 \, +\\
	&+
	\left(
		2 - R_0
	\right)
	\left[
		\bar{\sigma} ( 1 - \tilde\sigma ) +
		\frac{ \mu }{ \gamma }( 1 - \bar{\sigma} )
	\right] I^2 \, +\\
	&+
	\left\{
		\bar{\sigma}
		\left[
			R_0^{-1 } ( 1 - \tilde\sigma ) + \tilde\sigma
		\right] +
		\frac{ \mu }{ \gamma }
		\left[
			R_0^{-1 } ( 1 - \bar{\sigma} ) + \bar{\sigma}
		\right]
	\right\} I \, +\\
	&-
	\frac{ \mu }{ \gamma } \left( 1 - R_0^{-1 } \right) = 0 \; ,
\end{split}
\end{equation}
where $\tilde\sigma \defeq \sum_{i \in |\Lambda|} \sigma_{ij}^2 \; /(|\Lambda| \bar{\sigma})$.

For $R_0 >1$, eq.~\eqref{eq::I} always has a solution in $(0,1)$ which for $\mu / \gamma \ll 1$ 
is approximately given by :

\begin{equation}
\label{eq::I_approx}
I \approx \frac{\mu }{\gamma } \frac{R_0 - 1}{\bar{\sigma} [ 1 + \tilde \sigma (R_0 - 1) ]}  \; .
\end{equation}
This should be compared with $I = \frac{\mu }{\gamma } (R_0 - 1)$  for the prevalence of each strain of the classical SIR model in the same approximation, and shows how strain cross-immunity reduces overall prevalence.

In general, the polynomial $P_3$ satisfies $P_3(0)<0$, $P_3(1)>0$ and so it has either one root or three roots in $(0,1)$. For $R_0 < 2$, it is easy to check that there is only one root in $(0,1)$. For $R_0$ large enough, however, an additional pair of roots may exist for certain choices of $(\sigma_{ij})$. These two roots correspond to a stable and an unstable equilibrium for the total density of infected, and they are 'unphysical', in the sense that they have no correspondence with the SIR model and are a consequence of the closure assumption eq.~\eqref{eq::closure}. It can be seen that they are associated with cross immunity profiles which have values of
 $\sigma_{ij}$ well away from the values $0$ and $1$ for which the closure is exact. Moreover, the mechanism behind the high prevalence additional equilibrium is as follows: 
the susceptibility to reinfection is overestimated because cross-immunity is underestimated, and so infection of individuals who have been infected previously by other strains is overestimated.

\subsection{Agent-based model}

In order to check that this additional stable equilibrium is indeed an artifact of the closure assumption, we set up a fully stochastic agent-based model that directly implements the immune-history-based model as described by steps 1-5 of Section~\ref{S::Model} (retaining the assumption that any antibody against a strain confers total immunity). Taking parameter values for which the ODE model predicted a low prevalence equilibrium with $I \approx 4.3\e{-4}$ and $ \xi \approx0.93$, and a high prevalence equilibrium with $I \approx 0.66	$, $\xi \approx 0.27$, we performed simulations from initial configurations close to each of the predicted equilibria. 
The implementation of the initial configuration was made maximally random given those values as follows. First, a fraction $I / N_s$ of the population was infected with each of the possible strains. We considered a fully monoclonal population ($\sigma_{ij} = \sigma_{ij}^{(1)}$), which means that each infected agent carried exactly one antibody against the infecting strain. Then, we added a random antibody to the immune history of each agent, with uniform probability for each antibody, until the fraction immune to one arbitrary strain attained the equilibrium value for $\xi $. Assignment of a previously existing antibody was ignored. With this procedure, the resulting immunity to any strain is close to the equilibrium value.

\begin{figure}[ht]
\centering\includegraphics[width=0.8 \textwidth]{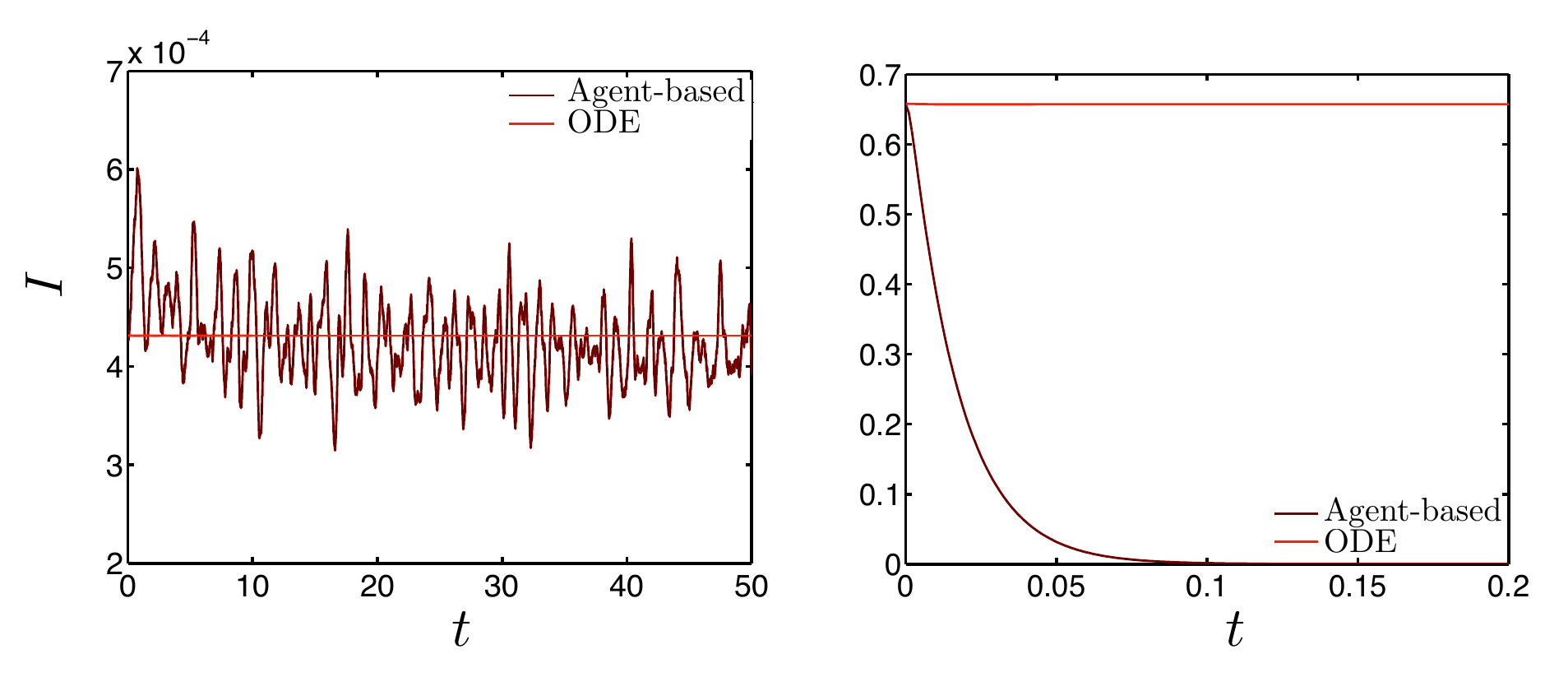}
\caption{Comparison of agent-based simulations with integration of the ODE model for the evolution
of the total prevalence with all strains circulating starting from \textbf{(left:)} initial conditions close to the high prevalence equilibrium, and \textbf{(right:)} initial conditions close to the low prevalence equilibrium. Parameters are: $n_e = 3$, $n_c = 2$, $\sigma_{ij} = \sigma_{ij}^{(1)}$, $R_0 = 15$, $\gamma = 90 \operatorname{year}^{-1}$ and $\mu = 70 \operatorname{year}^{-1}$. The agent-based simulations use $5\e{6}$ agents and are averaged over 30 runs.}
\label{Fi::multi_eq_sims}
\end{figure}

Simulations started off in this way for the values of $I$ and $\xi$ associated with the high prevalence equilibrium are shown in Fig.~\ref{Fi::multi_eq_sims}, right panel. In contrast, simulations started off for the low-prevalence values of $I$ and $\xi$ exhibit only fluctuations due to finite size effects, as one would expect in a stochastic realization of a true  mean field equilibrium, see Fig.~\ref{Fi::multi_eq_sims}, left panel.

\subsection{Degeneracy of the symmetric equilibrium and coexistence of non-symmetric equilibria}

We will now show how certain cross immunity profiles, which arise under common assumptions, can give rise to unexpected properties for the endemic equilibrium of a multi-strain system.

To see the relation between the properties of the cross-immunity matrix and equilibrium, assume that a symmetric equilibrium does exist. In particular, as we have just seen, this is always the case when all strains are circulating. Then, the equilibrium is degenerate if and only if:

\begin{equation}
\label{eq::degeneracy}
	\sum_{ j \in \Lambda } C_{ij} \delta_j = 0
\end{equation}
admits a nontrivial solution for $(\delta_j)_{j \in \Lambda}$ such that
$\sum_{ j \in \Lambda } \delta_j = 0$. Since by hypothesis the sum  $\sum_{ j \in \Lambda } C_{ij}$ must be independent of $i$ and the matrix $(C_{ij})$ is symmetric, the line sum  
$\sum_{ i \in \Lambda } C_{ij}$ is independent of $j$, and therefore any nontrivial solution $(\delta_j)_{j \in \Lambda}$ of equations~\eqref{eq::degeneracy} fulfills $\sum_{ j \in \Lambda } \delta_j = 0$. Since the sum of the components of any eigenvector associated with the zero eigenvalue of $(C_{ij})$ is null, any nontrivial element of the kernel of $(C_{ij})$ corresponds to a non-symmetric endemic equilibrium.

In general, $(C_{ij})$ is non-degenerate, but for some particular choices several properties of the cross-immunity matrix $(\sigma_{ij})$ contrive to produce a $(C_{ij})$ with a high dimensional kernel, as we will see shortly. 

Consider first the case when all strains are circulating. For most parameter choices, the matrix $(\sigma_{ij})$ is non-degenerate.
However, as discussed in Section~\ref{subS::strain-level}, a simple and natural choice for modeling cross-immunity is to take $\sigma_{ij} = \sigma_{ij}^{(1)}$, which corresponds to measuring the cross-immunity between two strains by the number of overlapping antibodies. The resulting cross-immunity matrices are directly related to the Hamming distance matrix, and this can be used to compute their spectrum \citep{gopalapillai}, which turns out to be highly degenerate and also quite simple: it has a positive (Perron-Frobenius) non-degenerate real eigenvalue, another positive eigenvalue with multiplicity $n_e (n_c-1)$, and a zero eigenvalue with multiplicity
$n_c^{n_e} - n_e (n_c -1) -1$.  The matrix $(\sigma^2_{ij})$ also has a positive (Perron-Frobenius) non-degenerate real eigenvalue, and, for $n_e \geq 3$ and $n_c \geq n_e-1$,
analysis of several particular cases reveals that it is also degenerate, with another two  
positive eigenvalues of multiplicities $n_e (n_c-1)$ and $n_e (n_e -1)(n_c-1)^2/2$, and a zero eigenvalue of multiplicity $n_c^{n_e} - n_e (n_c - 1)(1+ (n_e-1)(n_c-1)/2)-1$.
Also for this particular choice of $\sigma_{ij} = \sigma_{ij}^{(1)}$, analysis of numerous examples indicates that the eigenvectors of $(\sigma^2_{ij})$ are also eigenvectors of $(\sigma_{ij})$ and that, moreover, all eigenvectors of $(\sigma^2_{ij})$ associated with the zero eigenvalue are also eigenvectors of $(\sigma_{ij})$ associated with the same eigenvalue.

With closure assumption \eqref{eq::closure}, this implies the degeneracy of
the homogeneous system \eqref{eq::degeneracy}. Indeed, we can then write
$C_{ij} = {\tilde C}_{ij} + b_0 = b_2 (\sigma^2_{ij}) + b_1 ( \sigma_{ij}) + b_0$,
where $b_n$, $n=0,1,2$, are functions of $\xi $ and of the system's parameters.  
Equation~\eqref{eq::degeneracy} is equivalent to 
$\sum_{ j=1}^{N_s} {\tilde C_{ij}} \delta_j = 0$
because
$\sum_{ j=1}^{N_s} \delta_j = 0$.
The properties of $(\sigma_{ij})$ and of $(\sigma^2_{ij})$ stated above imply that $({\tilde C}_{ij})$ is degenerate with a zero eigenvalue with multiplicity
$n_c^{n_e} - n_e (n_c - 1)(1+ (n_e-1)(n_c-1)/2)-1$. Therefore the equilibrium value for the overall prevalence $I$ corresponds  in the $\eta_j^j$ simplex to a hyperplane of equilibria, given by the solution space of eq.~\eqref{eq::degeneracy}.  
The preceding argument holds only for the trivial distribution $p_1=1$, $p_{\{ 2 \leqslant \alpha \leqslant n_e \} }=0$, and depends on
the closure assumption \eqref{eq::closure} and on the presence of all strains.

\begin{figure}[ht]
\centering\includegraphics[width=0.4 \textwidth]{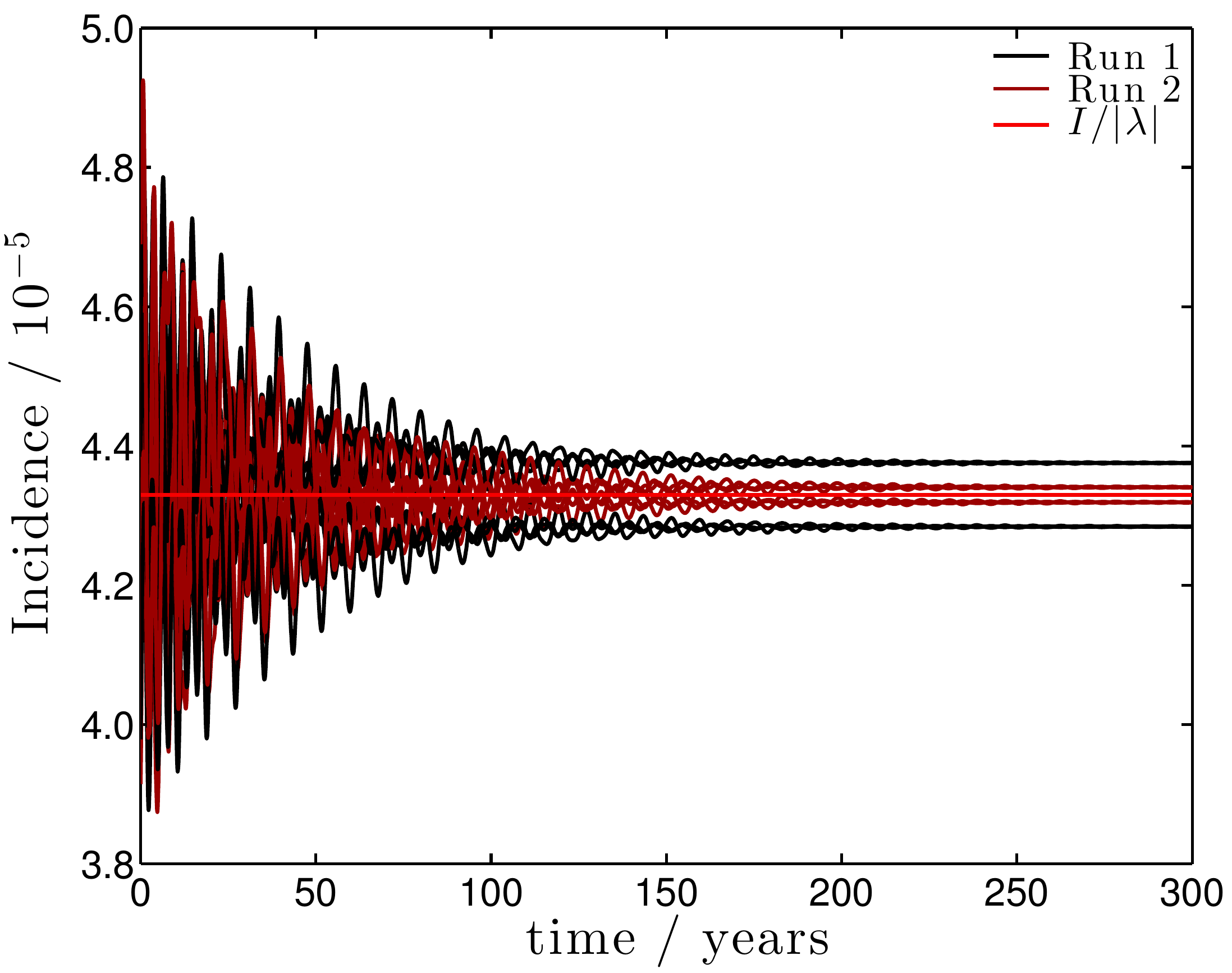}
\caption{Illustration of the degeneracy of the symmetric equilibrium of the ODE model. The incidence of each strain is shown for two runs with different initial incidences close to equilibrium. Parameters are $n_e = 3$, $n_c = 2$,  $\sigma_{ij} = \sigma_{ij}^{(1)}$, $R_0 = 5$, $\gamma = 90 \operatorname{year}^{-1}$ and $\mu = 70 \operatorname{year}^{-1}$. The equilibrium value of $I / | \Lambda |$ is also shown. It was obtained from eq.~\eqref{eq::I} and was verified to agree with that found for both runs.}
\label{Fi::deg}
\end{figure}

In Figure~\ref{Fi::deg} we illustrate this degeneracy effect for 8 strains, with $n_e = 3$ 
and $n_c = 2$. The plot shows orbits with different initial conditions converging to equilibria where, despite the symmetry of the system with respect to strain permutation, different strains have different prevalences. In this case, the dimension of the hyperplane of equilibria is 1. In the presence of mutations, the system may drift within this hyperplane of equilibria.

\begin{figure}[ht]
\centering\includegraphics[width=0.7 \textwidth]{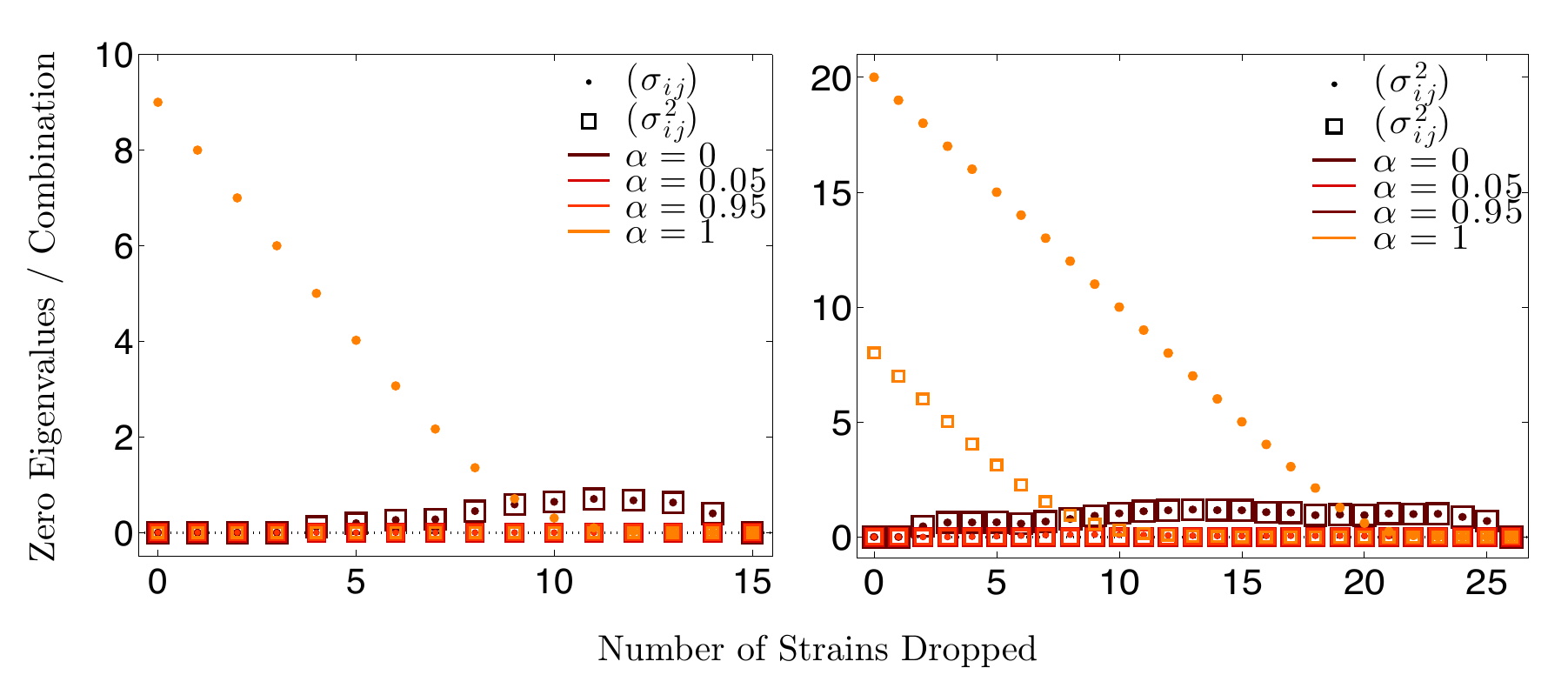}
\caption{Illustration of the degeneracy of $(\sigma_{ij})$ (dots) and $(\sigma_{ij}^2)$ (squares) for $\sigma_{ij} = \alpha \sigma^{(1)}_{ij} + (1-\alpha) \sigma^{(n_e)}_{ij}$, for different values of $\alpha$ and two choices of $n_e$ and $n_c$. (see main text for details).
\textbf{(left):} $n_e = 2$, $n_c = 4$.
\textbf{(right):} $n_e = n_c = 3$.}
\label{Fi::sigma_deg}
\end{figure}

To illustrate the behavior of the degeneracy of the cross-immunity matrix when an arbitrary subset of the possible strains is circulating, Fig.~\ref{Fi::sigma_deg} shows a measure of the number of zero eigenvalues of $(\sigma_{ij})$ and $(\sigma^2_{ij})$ with $i,j$ in each of the possible subsets. For each possible number of strains $N_d$ considered to not be circulating, let $C_{N_d}$ be the set of all possible combinations of circulating strains. Then the total number of zero eigenvalues for all possible combinations in $C_{N_d}$ is shown normalized by $| C_{N_d} |$. Three simple choices of cross-immunity profiles are shown for different choices of $n_e$ and $n_c$. We see that degeneracy is overall rare, although it is particularly high for the $\sigma_{ij}^{(1)}$ case mentioned above when most strains are circulating.

Closures involving higher powers of  $\sigma_{ij}$ will typically lift the degeneracy.
Writing $\xi_{ij}$ as a Taylor expansion in powers of $\sigma_{ij}$, we have, in equilibrium:

\begin{equation}
	\xi_{ij} = \sum_{n = 0}^N a_n( \xi ) \sigma_{ij}^n \; .
\end{equation}
The highest power $N$ can in principle be infinity, but is typically small. Substituting in the definition of $C_{ij}$ we have:

\begin{equation}
	C_{ij} = \sum_{n=0}^{N+1} b_n( \xi ) \sigma_{ij}^n  \; ,
\end{equation}
where:

\begin{equation}
\begin{aligned}
	b_n( \xi ) = a_{n-1}( \xi ) + \frac{\mu}{\gamma} a_{n}( \xi ) + \delta_{n1} ( R_0^{-1} - \xi ) \; , 
\end{aligned}
\end{equation}
with the convention that $a_{N+1} = a_{-1} = 0$. Using eq.~\eqref{eq::degeneracy}, we find the condition for equilibrium:

\begin{equation}
\label{eq::deg}
	\sum_{n=1}^{N+1} b_n( \xi ) \sum_{ j \in \Lambda } \sigma_{ij}^n \delta_j = 0 \; ,
\end{equation}
where again we have used the fact that $\sum_{j \in \Lambda} \delta_j = 0$ to drop the zeroth-power terms. The properties that relate the spectra and eigenspaces of $(\sigma^2_{ij})$ and $(\sigma_{ij})$ for $\sigma_{ij}= \sigma_{ij}^{(1)}$ do not seem to carry over to higher powers, and so for $N \geqslant 2$ the system \eqref{eq::deg} becomes non-degenerate even for the simplest cross-immunity profile $\sigma_{ij}^{(1)}$.

A common approach in the literature is to study the properties of equilibrium for simple models such as the one presented here \citep{ferguson:2004, ferguson:2005, ferguson:2008, cobey:2011}, and the particular cross-immunity profile associated with degeneracy is often chosen \citep{zinder:2013}. As such, we believe it is important to keep in mind that simple models with reasonable assumptions may lead to the presence of degeneracy; this degeneracy should be treated as unphysical, since, as can be seen from the discussion presented here, assuming other forms of cross-immunity, or most importantly increasing the order of the closure or assuming not all possible strains are circulating, will in general lift the degeneracy.

\section{Injection of mutant strains}
\label{S::Injection}

In this section we explore the behavior of our multi-strain model from 
an evolutionary perspective. Instead of assuming the presence of a certain number
of coexisting strains, as one is lead to by the analysis of the equilibria of
eq.~\eqref{eq::strains} in the absence of mutations, we shall investigate how 
diversity may develop from a single founding strain. We consider 
eq.~\eqref{eq::strains} for two strains, the founding strain 1 and strain 2 in its 
mutational neighborhood. We also invoke the closure assumption \eqref{eq::closure}.
For a particular choice of the distribution $p_{\alpha}$, $\alpha = 1, ..., n_e$, 
we assume this system to be in equilibrium with strain 2 absent, with the corresponding
single strain
equilibrium values $\xi _i$ and $\eta_{i}^j$, $i,j=1,2$, when a mutation occurs. 
To represent a mutation event in the ODEs framework we use, we 
introduce an implicit population size $N$ and 
use the initial conditions $\eta_{1}^1(0) = \eta_{1}^1 - 1/N$, $\eta_{2}^2(0) = 
\eta_{1}^2(0) = 1/N$,
with the remaining variables unchanged at their single strain equilibrium values.

\begin{figure}[ht]
\centering\includegraphics[width=0.4 \textwidth]{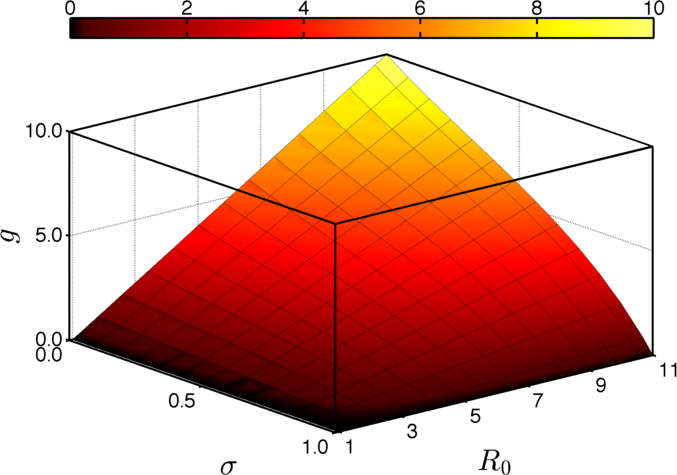}
\caption{Plot of the approximate initial growth factor $g$ of the invading strain as a function of $\sigma$ and $R_0$.}
\label{Fi::g_plot}
\end{figure}

In contrast with other models where invasion of escape mutants may
be conditioned, mutant strain 2 is always a successful invader. It will always
undergo a period of exponential growth, because immunity in the population is
below its equilibrium value. More precisely, it can be seen from 
eq.~\eqref{eq::strains} that the initial rate of this
exponential growth of $\eta_2^2 (\tau)$, in terms of nondimensional time $\tau = t/(\gamma + \mu)$, is $g = R_0 (\xi_1 \frac{\gamma + \mu}{\gamma} - 
\xi_{2} - \eta_{1}^1) \approx R_0 (\xi_1 - \xi_{2})$, where the approximation holds for $\mu \ll \gamma$. This growth factor is always positive.
In Fig.~\ref{Fi::g_plot} we show a plot of $g$ as a function of $R_0$ and the single non trivial
cross-immunity parameter $\sigma$ using the approximation:

\begin{equation}
\label{eq::g_approx}
	g(R_0, \sigma) =
		(1 - \sigma) (R_0 - 1)
			\frac{ \sigma (R_0 - 1) + R_0 }
				{ \sigma (1 - \sigma)(R_0 - 1) + R_0 } \;,
\end{equation}
which holds for $\mu \ll \gamma$.

We are interested in monitoring the behavior of the system beyond this 
initial phase of exponential growth, and check if any of the variables
$\eta_{1}^1$, $\eta_{2}^2$ ever goes below $1/N$, the value that corresponds
to a single individual in a population of size $N$ and therefore to extinction
of the corresponding strain. 

Even for a population size as large as $10^8$, the most common outcome of mutation implemented in this way is extinction of the founding strain, followed by extinction of strain 2. 
The reservoir of susceptibility to strain 2 in the initial population fuels a huge epidemic of
2, causing first the extinction of strain 1 and then that of strain 2 because of
a depletion of susceptibles, whose time scale for reposition is much slower.

\begin{figure}[ht]
\centering\includegraphics[width=0.5\textwidth]{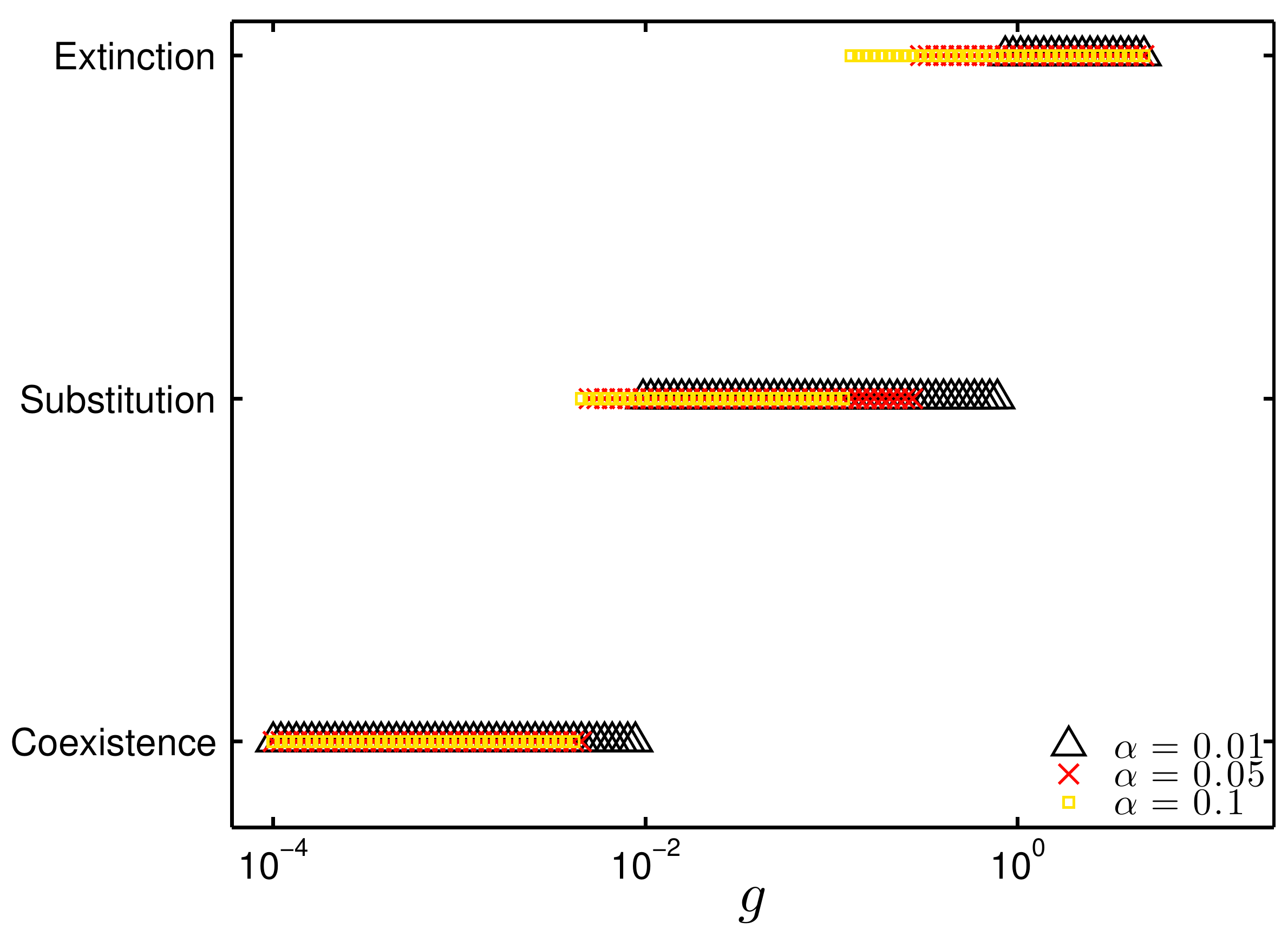}
\caption{lllustration of the different possible outcomes of an invading mutation occurring at the single strain equilibrium, as a function of the initial growth factor $g$ of the invading strain, for different values of $\alpha$. Note the qualitatively similar behavior regardless of the value of $\alpha$, whose specific value appears only to set the precise transition value of $g$ between different outcomes. Here, population size is $10^8$.}
\label{Fi::g_bif_1e8}
\end{figure}

This phenomenon is determined mainly by the rate $g$. Different outcomes can be engineered by fine-tuning parameter $g$ through the
choice either of  $R_0$ or $\sigma $. For small values of $g$, corresponding to $R_0$ close to one or to almost total cross-immunity, we may observe strain substitution, or even strain coexistence, after mutation (see Fig.~\ref{Fi::g_bif_1e8}). However, for these values of $g$ invasion by strain 2 becomes so slow that stochastic extinction events prevent these outcomes from being observed in agent-based simulations and most likely in real systems.

Note that relaxing the assumption that the original strain has reached equilibrium when the mutant is injected further favors extinction. A more detailed discussion of the implications of these results is presented in Section~\ref{Discussion}.

\section{Discussion and Conclusions}
\label{Discussion}

This work deals with multi-strain systems competing through cross-immunity
in an SIR type framework with vital dynamics. Each strain is characterized by a particular configuration, out of $n_c$ possibilities, of each of its
$n_e$ epitopes, and all $N_s = n_c^{n_e}$ strains have the same epidemiological 
parameters. 
The immune response of the population is heterogeneous,
in the sense that the number of epitopes to which antibodies are produced upon infection 
may vary in the population. It is homogeneous in the sense that any immune repertoire with at least one 
matching antibody confers full protection against a challenging strain.

In the spirit of \citep{levin:2007}, we built a reduced deterministic model
via a closure assumption that postulates the form of strain immunity
cross-correlations. For this reduced model, we studied the properties of the endemic equilibrium in the presence of the whole set of strains and of arbitrary subsets of this  set. We obtain two results that highlight a word of caution against possible unphysical 
consequences of seemingly reasonable assumptions made in the scope of analytic models. 

First, we find, in a region of parameter space, a high prevalence endemic equilibrium
in addition to the SIR low prevalence equilibrium. Comparison with agent-based simulations
show that this additional equilibrium is an artifact of the closure assumption.
Second, we obtain conditions for the endemic equilibrium to be symmetric, reflecting the 
symmetry of the full system with regard to strain permutation. 
We then find that for a particular choice of the immune
response profile of the population this symmetry is broken, and there exists
a manifold of non-symmetric endemic equilibria.  

The main goal is to explore the consequences of the basic assumptions made about the 
cross-immunity competition mechanisms in the presence of mutations. In particular,
we investigated if the former are compatible with the build up of antigenic diversity
from a single founding strain. Although the reduced deterministic model developed here 
may include the representation of mutations, it is well known this treatment of
discrete mutation events is too unrealistic. Here we adopted a mixed approach in which 
a cut-off value for the fraction of the population in each class stands for an
implicit population size, and mutations are implemented by an instantaneous change
in the values of these variables translating the switch of an individual infected with a 
given strain to an individual infected with a mutant.  

Extensive numerical integrations of the model show that, except in a small region of 
parameter space that allows strain substitution or strain coexistence starting from a founding strain, a mutation event leads to disease extinction. These results are confirmed by agent-based simulations,
which show furthermore that, due to stochastic extinctions, the regimes of  strain substitution and 
addition are not observed in reasonably sized populations. 
This is in contrast with the development of limited and even explosive diversity reported in the literature 
for similar models \citep{ferguson:2003, koelle:2006, minayev:2009a, minayev:2009b}. The motivation for these models is to find the conditions that reproduce the characteristic
phylogenetic pattern of influenza A, and extinctions are avoided either by working with 
continuous density variables and very low, perhaps unphysical, cut-offs  \citep{minayev:2009a, minayev:2009b}, or by including immigration of infectives \citep{ferguson:2003, koelle:2006}. The qualitative features of influenza A evolution are obtained when this is combined
with assumptions on strain space structure  \citep{ferguson:2003, koelle:2006} or on immune response  \citep{ferguson:2003, minayev:2009a, minayev:2009b}.

Apart from underlying SIR type dynamics with the same epidemiological parameters for all 
strains, the only ingredient of the model that was kept unchanged in our study is the 
hypothesis that, across the whole population, a single matching antibody confers immunity 
to a challenging  strain.
Given that the agent-based simulations are free of the limitations
of the reduced deterministic model and validate its predictions,
our results taken together strongly suggest that the build up of antigenic diversity 
from a single founding strain by invasion of mutants competing through cross-immunity
is incompatible with that assumption, in the framework of uniform SIR dynamics and
unstructured strain space. This conclusion is in agreement with the results
of a study based purely on simulations in which strain diversity is achieved by 
modifying  that hypothesis \citep{parisi:2013}. 

It is interesting to compare the results above with the fate of an antigenically identical
mutant that competes through increased infectiousness. In that case, strain substitution
is easily observed (results not shown). The model we develop here thus reveals a 
mechanism acting at the population level that would favor a serologically monotypic
virus evolving towards increased infectiousness.  For example, the virus of measles, a highly contagious disease, exists as a single serotype, despite having in vitro mutation rates  similar to  other viruses such as influenza that exhibit
large antigenic diversity. The measles virus also seems to conform to the hypothesis
of immunity being conferred by a single matching anti-body. 
It has been shown that successful escape mutants may require changes in all the epitopes 
of its hemagglutinin protein targeted by human antibodies, and that this 
plays a role in keeping measles serologically monotypic \citep{lech:2013}.

We speculate that the model developed here adequately describes the evolutionary epidemiology of a virus like measles, and that rapidly evolving, diverse pathogens such as influenza are associated with a weaker anti-body protection in some hosts.

\section*{Acknowledgements}

Two of the authors (TA and AN) gratefully acknowledge the support of Funda\c c\~ao Calouste Gulbenkian through its Programa Est\' \i mulo \`a Investiga\c c\~ao 2011, and of Funda\c c\~ao para a Ci\^encia e a Tecnologia through grants  PTDC/SAU-EPI/112179/2009 and PEst-OE/FIS/ UI0261/2011.

\bibliography{gripe}

\end{document}